# TWO EARLY-STAGE INVERSE POWER-LAW RELAXATIONS IN THE FAR-FROM-EQUILIBRIUM DYNAMICS IN SEMI-CLASSICAL PERCOLATIVE COMPOSITES

**Somnath Bhattacharya[1], Partha Pratim Roy[2] and Asok K. Sen[1]**
[1]*Theoretical Condensed Matter Physics Division; Saha Institute of Nuclear Physics*
*1/AF, Bidhan Nagar; Kolkata 700 064; India*
[2]*South Point High School; 82/7A, Ballygaunj Place; Kolkata-700 019; India*

**Abstract.** In several experiments for measuring various classes of responses, performed at least some four decades ago, on *driven* physical systems in a *far-from-equilibrium* (*or, from a steady-state*) *s*ituati*on*, early stage inverse-power-law relaxation dynamics had been observed. Since then, this intriguing behavior raised its head off and on until it regained its central role in the mainstream physical sciences about a decade ago with a *breakdown* and/or *avalanche* type (also called *self-organized critical*) behavior of the sand-pile model and a host of other similar problems. In this communication, we report on the non-equilibrium dynamics in our **Random Resistor cum Tunneling-bond Network** (**RRTN**) model. Previously, this semi-classical, or semi-quantum percolative model has been highly successful in explaining the static behavior for various random composite systems. In our dynamic studies for the last several years, we observe two initial power-laws (more than a decade each) and then an exponential relaxation for *asymptotically large time scales*. Efforts were made to interpret our results with various existing theoretical wisdom/s (which give, only one power-law relaxation for each such system near its breakdown or run-away type state). Obviously, our results (with two different power-laws) are richer than those particular cases. Further, a complete theory is still lacking probably due to a much deeper issue of entropy at stake. The appearance of two power-laws seems to be connected to some *non-extensive information-loss / entropy* (the experimental systems being mostly *athermal*) for such systems near their brinks (*catastrophic failure* not necessarily due to *criticality*).

## 1. Introduction:

Interest in relaxation phenomena in many diverse types of systems (magnetic, electrical, thermal, mechanical, fluid-dynamical, geological, biological etc.), has had a history of extensive research for the last four to five decades by physical scientists and engineers; and yet many experiments on such systems performed during the last decade, remain intriguing and mostly unexplained. Even the results of various theoretical models to explain these phenomena seem to be quite different due to the difference in the assumptions/inputs inherent in their formulations. The most important observation in the last decade regarding this issue has been an ubiquitous presence of early-stage inverse power-law relaxation of correlations (e.g., response functions, or their order-parameters) in such systems.

Of late, such systems have variously been called *soft condensed matter, driven, complex,* or *self-organized critical* systems and they force the theorists to delve deeper into their origins. Many attempts have been devoted towards finding out if any unified behavior exists among them [*e.g*., Refs. 1-6]. For example, the **dielectric breakdown** in random electrical composites made of conducting and insulating phases, and driven by a *dc* bias[1] ('*direct current*', fixed for one set of measurements at a time, across the opposite ends of a particular random sample, or configuration) had been studied in different experiments and there are several theories searching for an *universal feature* (or, *hidden symmetry*) of time-scale invariance in transport and relaxation in them (see ref. [2] and the citations therein, for many early attempts). So far, we were mostly interested in the *steady-state* (a time-independent *non-equilibrium state*, since it is an *open and driven system*) behavior, and a nonlinear *ac* (*alternating current*) *rms* (root mean square) response over half a period. To this end, the development of a semi-classical or, a semi-quantum percolation-based model, was undertaken in our group since the early 1990's and some early results were presented in the Ref. [6]. We call it a *Random Resistor cum Tunneling-bond Network* (**RRTN**) model (see Sec. 1.2 for some details). Further, new and detailed results on the *very low percolation threshold* plus some critical exponents using *finite-size scaling analysis* [7], the *nonlinear dc* characteristics [8], the *nonlinear ac* characteristics [9], and the *dielectric breakdown* [10], in the RRTN followed. The goal we are pursuing for the last three to four years has been investigating the similarities (referred to as a **time-scale invariance** in the Ref. [2]) of the relaxation dynamics observed in experiments mostly over the last decade, on a huge variety of systems, using the RRTN. In the present work, we report on our initial study on the dynamics and discuss some preliminary results on the appearance of two very distinct (inverse) power-law relaxations (implying *the absence of a time-scale*) in the early stage/s and their asymptotic crossover to an exponential relaxation (meaning the *evolution of a time-scale*) in the latest stage. These observations tend to establish that our model can not only recover the results of some earlier studies (both analytical and numerical), but it can also produce results beyond those achievable from any of the existing models/ theories. Indeed, it is gratifying for us to find that a simple RRTN model, originally proposed by us to explain *solely* the *ac and dc nonlinear response*, should show such a dexterity as to explain some other steady-state properties, e.g., a class of dielectric breakdown [10] as well as many dynamical behaviors like the relaxation dynamics in a *driven, far-from-equilibrium relaxation dynamics of conductance* [11], in a pleasantly unexpected fashion. In this short communication, we present the results of carrying out the *microscopic "Kirchoff's law-dynamics"* (basically enforcing the current conservation) at each node of the lattice, leading to two early-stage inverse power-law dynamics (before the asymptotic steady-state) in the RRTN model.

---

[1] It may be noted that the experimentalists prefer to work with a current bias, and the theorists with a voltage bias. But, whatever is the **chosen bias**, the response functions must behave **identically** for them.

**1.1 Percolation Models of Geometrical Connectivity:**

In English, the word *'percolation'* has a very widespread use since it is associated with almost everyone's daily life. We give here a brief, but self-contained review for our purpose; and for an uninitiated reader or one wishing to learn the subject actively, we recommend a relatively recent, but highly readable advanced text-book (with some very *useful computer programs* in the Fortran language), by Stauffer and Aharony [12]. It generally implies a mechanism of flow of some material (obviously, fluid-like) from one end to another of a solid-like (or, spongy/ porous) material filled with semi-microscopic obstructions. If it is used to model the flow of water through the narrow, short randomly placed pores of a *granular material* (or that of an electric charge through a *randomly dispersed metal* in an insulating matrix), one can imagine that the water (or, the current) will have to follow some tenuous path during its journey from one end of the material to the opposite one, along a *connected pathway* (a *system-spanning cluster*) if it exists. This purely *geometrically* motivated concept of the existence of an *abstract connected network*, under appropriate conditions (e.g., that of internally developing stress under an externally applied strain, in an elastic medium), through random multi-phase composites is known as *percolation* in the statistical physics of a microscopically inhomogeneous or granular medium, where the *inhomogeneity or the 'graininess'* appears at the semi-microscopic (much larger than atomic scale) level but the macroscopic system looks homogeneous.

There are several percolation models in the *statistical physics* literature. Basically, they are some simplified pictorial representations of its complicated real version. For example, there is a *continuum percolation* ( or, a *Swiss-cheese*) model of Halperin, Feng and Sen [13], where the inclusion phase inside a solid-like cheese-phase, is made of randomly placed *intersecting hollow* circles (in 2D) or *spheres* (in 3D). But, it is always simpler to understand the mechanism and easier to calculate (at least, numerically) the observable quantities, if the system under study possesses a discrete symmetrical geometry. So, one assumes the system to have an underlying lattice structure (e.g., a square lattice in 2D or a hyper-cubic lattice in a general *d*-dimension), with the *primitive lattice* size is chosen to be much smaller than the semi-microscopic size of the 'smallest' grains. Further, within the lattice models, there are sub-classes of bond or site-percolation models. So, if we are interested in the electrical transport in a random binary metal-insulator composite, then the host material may be considered to be an insulator and the metallic bonds or sites may be assumed to be randomly thrown guests. As the conducting bonds are distributed randomly in an insulating hypercubic lattice, this is a random resistor (or, conductor) percolation problem in that lattice. As a network, this problem in any dimension is called a *Random Resistor Network* (**RRN**) model.. For another class of composite materials, *e.g.*, a superconductor-metal (or, insulator) mixture, one may take the metal (or, insulator) phase to be the host and the random superconducting phases to be the guests.

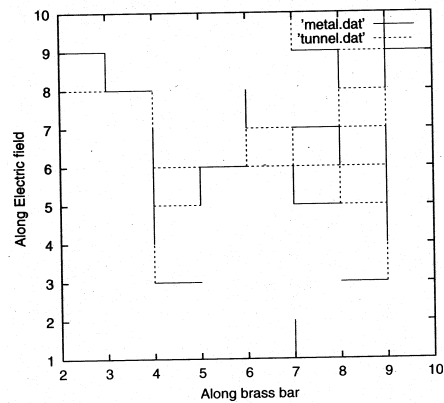

**Fig 1**. A 10x10 square lattice with randomly placed o-bonds (solid lines) only, at a concentration of p=0.15, in an insulating background (invisible blank bonds) is seen to be non-percolating. Even in the presence of all the possible tunneling bonds (RRTN with the t-bonds as dotted lines), it remains non-percolating or, insulating.

Further, it may be noted that there is *no explicit temperature* variable in any percolation model, and hence *no energy minimization principle* is available for the solution of the problem. But, in practice, some sort of **entropy maximum** due to statistical randomness or disorder in an **athermal system** (or, a thermal system at ***T =0***) should be involved, and may be defined through a loss of information content. So, if no properly defined entropy-maximization (or, *free-energy minimization*) principle exists, application of even a good/ rigorous mathematical procedure (e.g., as in the Refs. [4, 5]), may still lead one to an unphysical or, a partly correct, solution. In any case, it should be clear that for all the percolation models (exactly solvable or not), one deals with the '*geometrical connectivity*' between two opposite sides of a lattice. If a '*blind ant*' (technically, a *random walker*) is allowed to move from a position at one end of the sample, and it can reach the opposite end, by using *only* the conducting or ohmic (*o*) bonds (i.e., *the accessible paths* in a general physical system), then such a configuration is called **percolating**. Intuitively, it is easy to understand that the possibility of percolation increases with the concentration (*p*) of the *o*-bonds, present for a random configuration in any lattice. So, for a vanishingly low concentration (*p*) of the *o*-bonds, this connectivity is almost zero for any random configuration (i.e., a random sample). Also, it is obvious that at a very large *p* (close to 1.0), there would be *at least one connected pathway* (for a finite sized sample) of *o*-bonds from any one side to its opposite one. So, one expects a threshold value for this concentration, known as *percolation threshold* ($p_c$), where the connectivity, or a non-zero conductance, *on an average, is just established*. For a square lattice (with RRN), the bond percolation threshold is exactly $p_c = 0.5$. So this is the critical point for a *classical* (RRN) *metal-insulator phase transition.*.

One may also use other types of embedding lattices in the same Euclidean/ integral dimension, but then one typically finds that the values of only the so-called **non-universal quantities**, e.g., a critical point (*e.g.*, the Curie temperature, $T_c$, in a ferromagnetic transition problem, the threshold $p_c$ in a percolation problem, etc.), change. There is no rigorous proof but it has been inferred from a very large set of experience (e.g., with intuition, computer experiments, semi-rigorous arguments, etc.) that the nature of criticality (say, the power-law exponents), for various measurable scaling-functions in the domain of criticality, remain invariant of the type of lattice in any fixed dimension, *d*. This amazing property is called the **universality of the exponents**. Further, rigorous statistical analysis reveals that at $p_c = 0.5$ with a system-spanning cluster (the cluster of conducting bonds, which spans the opposite sides of the macroscopic sample), the system has a fractal, or a non-integral dimension. This fractal nature of the percolating system is generally responsible for the power-law behavior of the measurable quantities at the *classical* percolation threshold (or, $p = p_c$) *only*.

Before ending this brief review, let us also remark that there also a class of *correlated percolation* problems, where the occupation probability of a given species at one position is *statistically correlated*, in some chosen semi-random fashion, to that at some other neighboring site/s. Mathematically exact solutions (taking care of the fluctuations, neglected in the mean-field methods) for most of the RRN's, using stochastic *differential* (for *continuum* models), or *difference* (for *lattice* models) *equations* with '*white noise*,' particularly above 2D, is difficult to come by. For obvious reasons, a correlated percolation problem is, in general, even more difficult to solve exactly mathematically (except for some correlations, custom-made with hindsight), since one needs to solve for the stochastic equations with a so-called '*colored noise*' for the randomness. Hence, for more realistic correlations, one mainly takes recourse to either some analytical mean-field (*e.g.*, effective medium) *approximation*, *i.e.*, EMA, or some '*exact*' computational method. Indeed, these are what we would also have to do for the RRTN model (see below).

## 1.2 Random Resistor cum Tunneling-bond Network (RRTN) model:

To capture the basic physics for those random metal-insulator mixtures with very low percolation threshold, a lattice-based bond-correlated percolation model was formulated [6-8]. There are many dc and ac transport measurements in metal-insulator type composites, like Carbon-black-PVC (or, some other polymers), which show a finite response at *p* much below their $p_c$, the percolation threshold of their doped metallic phases. So, to explain these and some other observations (see Refs. [6], or [8], for a comprehensive list), in particular the *low temperature resistance minimum of the bulk insulators*, it was necessary for us to introduce a scope of conduction beyond all the classical possibilities, which had already been taken care of in the RRN-type models described above. For this purpose, in addition to the ohmic transport in the RRN, a *tunneling* process was introduced in a *semi-classical* or *semi-quantum* fashion. It is semi-quantum because only tunneling through the energy barrier due to an insulating (dielectric) bond between **any two nearest neighbor ohmic bonds (o-bonds) only** is allowed by modeling a *microscopic energy-gap* (or, an equivalent *voltage threshold*, $v_g$) between them, while disregarding any quantum mechanical phase-information of a charge carrier passing through them. We call an insulating or, a *gap* bond, which becomes a *bridge* at or above a potential difference of $v_g$, a tunneling (*t*) bond. If the underlying RRN in a particular RRTN configuration (or, sample) is non-percolating, then that macroscopic RRTN will also have a *macroscopic threshold voltage* $V_g$, below which there is no response (no current at any finite macroscopic voltage, $V < V_g$).

The placement of *only the t-bonds* in a *deterministic* fashion, *statistically fully correlated* with the *random positions of the o-bonds* in any particular configuration, has made our RRTN model partly deterministic, or *conditionally t-bond correlated*. We have calculated the percolation threshold of our RRTN model [8, 9] in the presence of all the possible *t-bonds*, *i.e.*, in the *asymptotically high voltage regime* with only the reversible tunneling and *no irreversible* Joule heating, and named it $p_{ct}$, to distinguish it from $p_c$, by adding the extra subscript '*t*' for this *tunneling percolation*. In a 2D square lattice RRTN, using finite-size scaling analysis, it comes out to be $p_{ct}=0.181$ [7], whereas the mean-field (EMA) value for $p_{ct}$ is 0.25 [8]. If the bulk *dc* differential conductance $G(V)$ ($=dI/dV$) vs. *V*, the external driving voltage, is studied for the RRTN model, one finds a *sigmoidal* type curve, *highly asymmetric* in the higher voltage regime. This high asymmetry matches with the experiments also, and has important bearings on our mathematical expression for the response function.

If the underlying RRN is insulating, then at sufficiently low driving voltages ($V < V_g$), no *t*-bond is active, and the RRTN is also insulating. But, if the *o*-bonds are percolating and the driving *V* is quite low, then no *t*-bond is able to enhance *G* which remains a constant independent of *V* and so the $G(V)$ *vs. V* graph is horizontal to the *V*-axis. This region is termed as the *lower linear regime*, and the voltage regime close to $V_g$ is called, the asymptotic lower voltage regime. The word *asymptotic* is used here because in highly structured and ultra low-conductance composites (even at a medium-low temperature of about $100^o$ K) with a *ubiquitous nonlinear response*, it is very difficult experimentally to even approach this regime, and obviously to find out the value of $p_{ct}$. Thereafter, under further increment of the external voltage *V*, more and more *t*-bonds will tend to form several parallel paths. Thus in this region the differential conductance will rise monotonically with voltage and the nonlinear response in $G(V)$ will start from this point. This is the *sigmoidal* region of the ***I-V*** characteristic. At a very high enough driving voltage (but, in the **absence of the irreversible Joule-heating effect, in principle**), all the allowed *t*-bonds are actively participating in current conduction and thus the increase in $G(V)$ with the increase in *V* is absent here. So, the $G(V)$ *vs. V* curve is parallel to the *V*-axis (ohmic) in this regime, called the *upper linear regime*.

The main claim of this particular discussion should thus impress that the threshold $p_{ct}$ of our RRTN model, with even the smallest distance (just one lattice constant) tunneling cut-off, in any embedding dimension and the type of lattice, but is also surprisingly smaller than that of the classical RRN model. If further neighbor tunneling is allowed, the $p_{ct}$ becomes even smaller in support of the experiments, *e.g.*, [ ] on Ref. [14] on the doped polyaniline). Further, the *random placement* of the *o*-bonds and the imposition of the *energy-gap* $v_g$ through the t-bonds in the RRTN model, imply that the *disorder* and *interaction* respectively are empirically built-in, in this model.

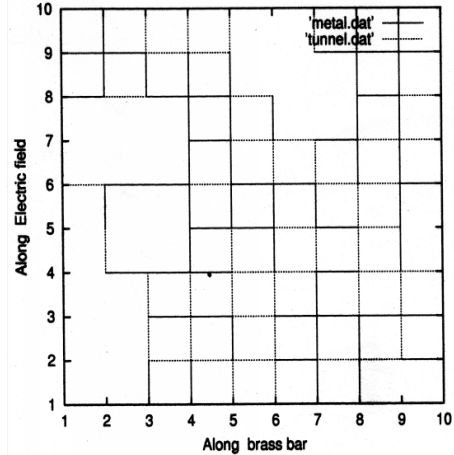

**Fig. 2**. Another 10x10 square lattice (RRN with the solid lines only as the random o-bonds) at a concentration of p=0.40 is found to be (classically) non-percolating. But, in the presence of all the possible t-bonds shown as dotted lines (i.e., a semi-classical RRTN configuration), at least, one system spanning cluster is found. So this configuration is classically non-percolating, but semi-classically (as a RRTN) percolating, or metallic.

## 2. Study of Relaxation:

### 2.1. Types of Relaxation Phenomena:

In the collection of *dc* transport data, one throws away the information of the early-stage (mostly, internal rearrangements) non-equilibrium dynamics of the sample towards its steady state. What we get is only the time-invariant (*average constant*, except for some *noise terms*) response of the system to the external driving field. Actually, no macroscopic system in nature responds instantly to an external perturbing field (the principle of '*action at a distance*'), but takes a finite time (may be very large, though) to '*fully*' respond to the situation. In the literature, there are *two* classes of relaxations. (i) *Debye-type*: It follows a pure exponential function in time, with a single time-constant. Debye, in his classic treatment of dielectric relaxation in a fluid, with a dilute concentration of dipolar molecules, had derived the relaxation function, $\phi(t) = \phi(0) \exp(-t/\tau)$, where $\tau$ is the *time-constant* (or, *relaxation time*, which is, obviously, defined as the time needed for reaching $(1/e)$-th fraction of the initial value); and (ii) *Non-Debye type*: This follows a linear superposition of several exponential functions with multiple time-constants (extended later on, to include a sub- or stretched-exponential function, $\phi(t)= \exp(-t/\tau)^\mu$ where $0< \mu <1$, which shows up in glassy systems). Of course, mother Nature may have in store, some other possible functions as well, where the system relaxes fine, but the rigorous definition of a time-scale, $\tau$, may fail. They are quite exotic and do *not possess any time-scale*, at all, or has all possible time-scales in them. Clearly, this behavior is non-Debye type and indeed, they used to be found in the earlier years, in the *long-time tails* of the relaxing entity. They used to be considered more as an exception than the rule. But, then many experiments started finding power-law decays (correlations) extremely close to any critical phase transition. Indeed, physicists started using the phrase "*critical slowing down*" for the approach to the critical point in the time-domain. But, the problem did not die down there; and during the last decade or more, people from almost all walks of science have been finding power-law relaxation/s in many realms. Being in a confused state in an empowering maze, scientists (or, science journalists and authors of popular science books) started calling it '*complexity physics*,' '*soft condensed matter*,' etc. With dusts settling, somewhat more thoughtful scientists compared this lack of any time scale with the critical slowing down, and named it as '*self-organized criticality* (SOC),' since in such cases were also observed in nature (as in earthquakes, etc.) for a long time, and mostly in some mathematical models in relatively recent times. Our viewpoint, to be demonstrated in the sequel, is that the main reason is the non-linear response of a measurable quantity, and the SOC is only a small part of the story in systems with linear response property (*e.g.*, the RRN model).

Coming back to the present problem, though there is a single external voltage across the sample, the internal field across each microscopic phase (*o*-, *t*- or *i*-bonds) can never be uniformly distributed, and each of them responds differently to the same amount of internal field across them. So we can say, in a mathematical sense, that a bulk sample possesses a distribution of relaxation times inside its microscopic domains of *micron*- and *nano-sized clusters*. That is why any relaxation study reflects the interaction and consequent evolution (though in a combined fashion) among all the microscopic clusters having different $\tau$'s. To study experimentally how the bulk macroscopic response of a particular sample depends on its microscopic parts, without changing the sample itself (*e.g.*, AFM, or the Atomic Force Microscopy, muddles the part of the sample examined, thoroughly) there is no other way than to study its relaxation dynamics. But, in this case, the experimentalist would have to use some theory or a theoretical model to extract the information/s on even the statistical properties of different types of clusters (not, to speak of the individual clusters at all!).

As discussed earlier, to achieve quantitative accuracy in the values of measurable entities in a real sample, it may be quite important an experimentalist to make an *ansatz* for a pre-conceived distribution (a continuous and normalized probability density function) of the relaxation times. For example, in the presence of a crossover from one (i.e., local) fractality to another one (*i.e.*, global/ macroscopic), Mandelbrot et. al. [15] find that a special class of non-Debye relaxation leads to the power-law relaxation dynamics. Further, inside a disordered sample, the response of some of the semi-microscopic grains or phases may also be partly nonlinear (*t*-bonds in the RRTN model). Also, the *microscopic voltage threshold* value ($v_g$) for the *charge transfer through* an insulator may possess a random probability density function. If not quite necessary for the basic physics under study, these extra attributes bring more empirical parameters in the theoretical study of relaxation for random (i.e., positional disorder) multi-phase composites unnecessarily complicated. As physicists, we are interested mostly in the similarities of the qualitative (if not quantitative, as for the critical exponents) aspects unifying a large variety of very different types of systems. As a result, we look for the basic origin of these similarities and hence keep our model/ theory as simple as possible, using the smallest number of empirical parameters. So, for our studies so far and in this paper, we have kept $v_g$ constant.

In addition to the general considerations of RRTN model, especially for studying the relaxation dynamics under a fixed *dc* voltage-bias, we have introduced the contribution of *displacement* current for the tunneling bonds (or *t*-bonds), when the voltage ($v$) across a *t*-bond is below its pre-assigned threshold value $v_g$. In other words, a *t*-bond behaves as a capacitor if $v < v_g$ and possesses a microscopic conductance g, equal to that of an *o*-bond if $v > v_g$. This assumption is physically acceptable due to the possibility of a dielectric breakdown of the insulating material, placed between two plates (or, between two ends of a *t*-bond), in a *schematic capacitor*.

## 2.2. Experimental Evidences for Two (inverse) Power-law Relaxations:

About 32 years ago in 1970, the intriguing relaxation phenomenon of two very distinct (inverse) power-law behaviors in the early time-stages (several decades though) in a specific photoconductor, namely, amorphous $As_2Se_3$, was reported [2]. A sufficiently good theory, based on a random-walk was given there, shows that the transient current grows with a sublinear function before some crossover time $t_r$ and then the profile follows a superlinear form with time (t). In a recent paper [3], Stadler and Mehta found at least two power-laws in their simulation with vibrating sand. In dense granular media [7], which behaves as *athermal* glass, the deviation from exponential relaxation is also prominent. Even in natural earthquake, people are getting the trace of more than one power-law dynamics. These systems are surely far from their equilibrium situation and all of them follow a purely exponential nature in the situation, very close to equilibrium. Non-exponential kinetics leading to a power law behavior is also found in CdSe quantum dots [5], as well as there is *two power-laws in the* growth kinetic study of sputter deposition [6] by Ag on Si/ $SiO_2$.

Weron and Jurlewitz [1] showed that $\tau(t)$ obey a *power-law time-series* with a crossover region between them. In some attempts, Naudts et al [4] discussed the scope of non-extensive thermo-statistics to explain the power-law dynamics in different samples.

### 2.3. Results:

In our numerical study, we have applied a fixed dc bias across the square lattice RRTN network of sufficient system size. We observed a non-Debye type relaxation behavior. Thereafter, to compare and analyze the transient currents from our numerical results as a decaying fashion (*each of them being about one decade or more in range with*), we have subtracted the fixed steady-state value from each of the time-dependent current. As result, in each current decay profile, the same power-laws were found to appear (i.e., the same as in their un-subtracted actual time-series).

We demonstrate this in Fig.3, where the graph shows that the first (inverse) power-law has a scaled time range of about $t$=20-200, with an associated exponent of $\alpha$=0.55. The second one of the two initial power laws (after a relatively small crossover time-region beyond the first one), appears in the scaled time range of about $t$=1500-20000, with the other associated exponent of $\beta$=0.826. It must also be mentioned here that for any $p$, or for any random configuration at the same nominal $p$; (i) the dynamics always shows a pair of $\alpha$ and $\beta$ (sometimes, though rarely, we find only one power-law), so that **more than two power-law dynamics is never observed**; (ii) there is always a finite cross-over time between these two power-law decays, so that there is **no critical phase-transition involved between these two initial dynamics**; and finally, (iii) there is always a finite cross-over time domain between the second power-law dynamics and the next one, which is always **a pure exponential decay** at the end. Thus, there is no phase transition in the final round either, and the exponential decay implies that the driven sample is approaching a steady-state, much like that in a simple CR (combined resistor-capacitor) circuit.

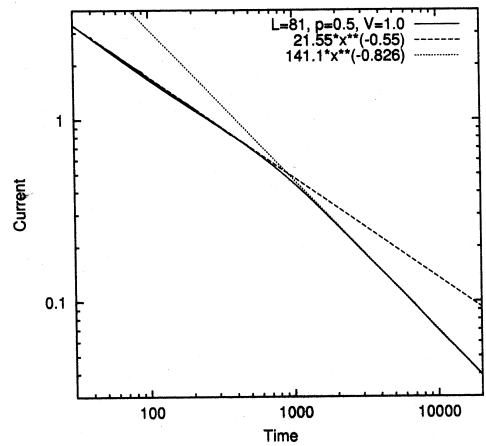

**Fig 3.** Relaxation dynamics of a percolating two-dimensional square lattice (2D RRTN), with L=80, p=0.5, and an external driving voltage of V=1.0. The two initial (inverse) power-law relaxations have the exponents $\alpha$=0.58 and $\beta$=0.83 respectively, with a small yet continuous crossover time-domain between them. Finally, the approach to its steady state takes through a purely (decaying) exponential tail.

We wish to reiterate that we obtain these two power laws (of course, both of them as a pair, being different for different for different p's) for *any concentration*, far from the classical critical concentration of $p_c$ = 0.5 (but greater than our semi-quantum critical concentration of $p_{ct}$=0.181 in the RRTN). This shows that the origin of these pair of power laws is ***not due to the closeness to any critical (phase transition) region, but due to the fundamental reason of non-linearity inherent*** in the formulation of our RRTN model (see the *Discussion section* below). Further detailed work in this respect is in progress, and will be discussed elsewhere.

### 3. Discussion:

The appearance of these couple of power laws seems to be due to the dynamics of the microscopic structures (in real-space) evolving from local (fractal-like) clusters to a global (fractal or fractal-like) system-spanning cluster in the underlying lattice [6] (also the ref. [10]). As an external voltage *V* starts driving the macroscopic network, the microscopic voltage (*v*) distributions at each site of the network start changing, and so does the voltage difference across each bond. But, due to their unequal response-characteristics (i.e. the conductance of the conducting, insulating and the coherently (i.e., in a perfectly statistically correlated fashion) positioned tunneling bonds, for the same microscopic field, the microscopic relaxation initially takes place in each local cluster almost separately. But, because of the mismatch in the microscopic conductance between the nearest neighbor bonds, the competitive efforts to bring about a global steady state are initiated within the local environments, where both the *o*- and the *t*-bonds have important roles to play. That leads towards a goal of achieving cluster-wise locally steady- states. Obviously, before these efforts are concluded, the intra-cluster interactions (through tunneling only) for the next level of efforts towards a time-dynamic steady state with a significantly different power-law. Both the dynamics for different length scales (i.e., inter and intra) have their own fractal-like nature, even quite far from critical concentration and they bring two different power laws in the bulk, or macroscopic, dynamical behavior. With a different driving field (like ionic concentration differences in a chemical or biological system), it is found using a discrete, one-dimensional, stochastic equations for the *Calcium-channel dynamics in living cells* [15], we observe again that the relaxation follows two early, very distinct power-laws with a crossover for about four decades ($10^1$-$10^5$ seconds). Clearly, this property has a close similarity with that of our RRTN model (though the RRTN is used in 2D, as in this paper, and their stochastic model is in 1D).. Even though the authors of [15] did not mention this intriguing issue since they had different message/s to convey in that work, the existence of two power-law relaxations in the calcium channels of living cells was highly appreciated [15(b)].

Finally, the consideration of *nonextensive* generalized entropy, such as that of Ren'yi, Shannon, or a combined one due to Tsallis etc., leads to some early power-law dynamics and is well discussed in some recent literature ( **e.g**., [4, 5]} and many more). Work in this direction is under progress.

## 4. Acknowledgements:

AKS is thankful for the warm hospitality and support of Profs. P. Sheng and P.M. Hui at an International Conference named ***ETOPIM5***, June 1999 (see [9] below), where early results on relaxation in the RRTN were presented. Likewise, SB is thankful for the warm support of the organizers of the Workshop ***SMR1322*** (year 2001) and ***SMR1519*** (year 2003), held at the Abdus Salam International Centre for Theoretical Physics (AS-ICTP), where further progress was presented. SB acknowledges useful discussions with the SAND group of the AS-ICTP, on the relationship of this work with those on some earthquake studies. AKS acknowledges very fruitful discussions with Prof. A. Mehta on the analogy of these two power-law relaxation of the RRTN model with rather recent similar observations in some *real earthquakes*; and with Prof. H. Levine on their discrete one-dimensional stochastic model of *Calcium-channel dynamics in living cells* [15]. Further, both SB and AKS gratefully acknowledge the warm hospitality and gracious support of the convenors (Profs. G.D. Roy and M.A. Hossain) and the organizers of an International Conference on Mathematics and Mathematical Physics (***ICAMMP 2002***) held at the Dept. of Math. of the Shahjalal Univ., Sylhet, Bangladesh, where SB was awarded a gold medal, by an international panel of judges, as one of the best presentations (among the Ph.D. students and Post-docs). Again, we are thankful to the AS-ICTP, since it was one of the sponsors of the ***ICAMMP 2002***.